\documentclass[conference]{rss}
\usepackage{listings}
\usepackage{times}

\usepackage[numbers]{natbib}
\usepackage{multicol}
\usepackage{amsmath}
\usepackage{graphicx}
\usepackage{gensymb}
\usepackage{amsmath, xparse}
\usepackage{mathrsfs}
\usepackage{soul}
\usepackage{float}
\usepackage{xcolor}
\usepackage{stfloats}

\usepackage[bookmarks=true]{hyperref}

\definecolor{linkblue}{RGB}{40, 0, 255}

\def\BibTeX{{\rm B\kern-.05em{\sc i\kern-.025em b}\kern-.08em
    T\kern-.1667em\lower.7ex\hbox{E}\kern-.125emX}}
\newcommand{\figref}[1]{\textcolor{linkblue}{Figure~\ref{#1}}}

\newcommand{\secref}[1]{\textcolor{linkblue}{Section~\ref{#1}}}

\newcommand{\appref}[1]{\textcolor{linkblue}{Appendix~\ref{#1}}}
\def\BibTeX{{\rm B\kern-.05em{\sc i\kern-.025em b}\kern-.08em
    T\kern-.1667em\lower.7ex\hbox{E}\kern-.125emX}}

\begin{document}

\title{Effect of Adaptive Communication Support on LLM-powered Human-Robot Collaboration}


\author{Shipeng Liu$\mathsection$, FNU Shrutika$\ast$, Boshen Zhang$\ast$, Zhehui Huang$\ast$, Gaurav Sukhatme$\ast$,and Feifei Qian$\mathsection \ddagger$\\
 $\ddagger$ Corresponding Author\\
$\mathsection$ Ming Hsieh Department of Electrical and Computer Engineering \\$\ast$ Thomas Lord Department of Computer Science\\
University of Southern California\\
Los Angeles, California, USA\\
\{shipengl,sshrutik,boshenzh,zhehuihu, feifeiqi\}@usc.edu
}


\maketitle

\begin{abstract}
Effective human-robot collaboration requires robot to adopt their roles and levels of support based on human needs, task requirements, and complexity. Traditional human-robot teaming often relies on a pre-determined robot communication scheme, restricting teamwork adaptability in complex tasks. Leveraging strong communication capabilities of Large Language Models (LLMs), we propose a \textbf{H}uman-\textbf{R}obot \textbf{T}eaming Framework with \textbf{M}ulti-Modal \textbf{L}anguage feedback (HRT-ML), a framework designed to enhance human-robot interaction by adjusting the frequency and content of language-based feedback. HRT-ML framework includes two core modules: a \emph{Coordinator} for high-level, low-frequency strategic guidance, and a \emph{Manager} for subtask-specific, high-frequency instructions, enabling passive and active interactions with human teammates.
 To assess the impact of language feedback in collaborative scenarios, we conducted experiments in an enhanced Overcooked environment with varying levels of task complexity (easy, medium, hard) and feedback frequency (inactive, passive, active, superactive). Our results show that as task complexity increases relative to human capabilities, human teammates exhibited a stronger preference towards robotic agents that can offer frequent, proactive support. However, when task complexities exceed the LLM's capacity, noisy and inaccurate feedback from superactive robotic agents can instead hinder team performance, as it requires human teammates to increase their effort to interpret and respond to a large number of communications, with limited performance return. Our results offer a general principle for robotic agents to dynamically adjust their levels and frequencies of communications to work seamlessly with humans and achieve improved teaming performance.

\end{abstract}
\IEEEpeerreviewmaketitle

\section{Introduction}
Human-robot collaboration has been extensively studied and applied across diverse scenarios, demonstrating strong potential towards enhanced efficiency and performance~\citep{jahanmahin2022human,chuah2021future,park2020active,gordon2020adaptive,xiao2020three,Liu2023,10.1145/3610977.3635112}. As task complexity increases, robot adaptability becomes increasingly essential for seamless teamwork. Previous work has developed methods and tools for robot to adapt their actions based on inferred human objectives~\citep{10.1145/3610977.3635112}, trust level~\citep{chen2020trust}, and individual  preferences~\citep{biyik2022learning}. Recent work also begin to incorporate language-based feedback~\citep{ozdemir2022language,sharma2022correcting} to enable more direct communications and lower user barriers. However, robot communications in these approaches primarily focused on relative simple commands such as ``pick up the book'' or ``move left a little'', which do not reflect the level of human-robot communications required in real-world applications.  

Recent advancements in large language models (LLMs) have brought powerful reasoning~\citep{zhang2023proagent,zhang2024combo,agashe2023evaluating,guan2023efficient}, natural language understanding~\citep{liu2023llm,wuautogen}, contextual awareness~\citep{deng2023rethinking}, and generalization capabilities~\citep{ge2024openagi}, enabling more advanced, prolonged communications~\citep{bubeck2023sparks,ouyang2022training,hong2023metagpt}. These methods has empowered robots to process ambiguous and complex instructions from human~\citep{liu2023llm}, engage in more natural and dynamic conversations~\citep{wuautogen,hou2024my}, and learn from a diverse set of inputs~\citep{ge2024openagi,sun2024optimizing}.

However, even in these LLM-enhanced communications, human teammates continue to play a predominant role in requesting specific tasks and providing suggestions during collaboration~\citep{liu2023llm}. ~\citet{schoenegger2024ai}'s study suggested that state-of-the-art LLMs can often match or surpass human performance in various domains. Based on these results, we hypothesize that allowing LLM-powered robots to participate more proactively in~\citep {tanneberg2024help} or even initiate communications with human teammates can enhance teaming performance and efficiency. 

To test this hypothesis and systematically evaluate the impact of different forms of language feedback provided by robots on collaboration efficiency and human satisfaction, in this study 
we develop \textbf{HRT-ML}, a human-robot teaming framework incorporating multi-modal language feedback to support dynamic, context-aware teaming styles.
To enable the robot to provide effective language feedback, HRT-ML comprises two main modules: a \textit{Coordinator}, which manages overall collaboration strategies and delivers low-frequency or passive instructions and feedback, and a \textit{Manager}, which determines appropriate subtasks based on the coordinated plan at each stage, offering high-frequency instructions. Combining \textit{Coordinator} and \textit{Manager} allows the robotic agent to provide different forms of instruction at different frequencies. To investigate how the form and frequency of language feedback influence teaming performance, we performed user studies using four different robot active levels: Inactive, Passive, Active, and Superactive. We find that as the environment becomes more challenging, participants exhibit a stronger preference for robots that provide active support, whereas in simpler tasks, frequent feedback is often perceived negatively, reducing overall performance and human satisfaction. These findings reveal the importance for the active level of robot language support to dynamically adapt based on task complexity and team capability. Inaccurate or overly frequent feedback can decrease team efficiency, as participants must take extra time to understand and correct these suggestions. 

In summary, our contributions are as follows: 
\begin{itemize}
    \item developed a human-robot teaming framework incorporating multi-modal language feedback to support dynamic, context-aware teaming styles.
    \item performed user studies to determine how forms and frequency of robot language feedback influence teaming performance.
    \item discovered how the active level of robot language support should adapt based on task complexity and team capability, to best enhance teaming performance.
\end{itemize}
To the best of the authors' knowledge, this is the first study that systematically explores the effect of different types of LLM-based multimodal language feedback on teaming performance across different task complexities.

\section{Related Work}

Robots serving as teammates and collaborating with humans in shared workspaces is an emerging paradigm in human-robot interaction~\cite{christoforou2020overview, faccio2023human, zhang2024mutual}.
Recent advancements in LLMs have further enabled robots to interpret human intent, generate task plans, and engage in more natural interactions~\cite{zhang2023building, li2024conav, zhang2024mutual}.
In our work, we specifically focus on the role of communication in LLM-based human-robot collaboration, as effective information exchange is crucial for task coordination, intent alignment, and adaptive interaction.

Communication in human-robot collaboration can generally be categorized as either unidirectional or bidirectional.
In unidirectional communication, robots do not proactively engage with humans; instead, they infer human intent from instructions and make autonomous decisions based on observed cues~\cite{zhang2023proagent, chang2024partnr, li2024conav, liu2023llm}.
In contrast, bidirectional communication involves robots actively interacting with humans—seeking clarification, sharing updates, and negotiating task execution strategies~\cite{guan2023efficient, zhang2023building, lu2024proactive, zhang2024mutual}. 
Several studies have explored unidirectional and bidirectional communication strategies, finding that humans perceive robots capable of bidirectional communication as more intelligent and collaborative~\cite{ashktorab2021effects, zhang2024mutual}. 

However, existing studies~\cite{ashktorab2021effects, zhang2024mutual} on bidirectional communication primarily focus on updating status or requesting simple, predefined subtasks without incorporating complex reasoning about the coordination strategies. A well-defined bidirectional communication framework for complex coordination tasks is still lacking. Moreover, the impact of different forms and frequencies of bidirectional communication on coordination performance and human preference remains unexplored.

\section{Testbed: Overcooked-Environment}
\begin{figure}[ht]
\centering
\vspace{0.08in}
\includegraphics[width=1\linewidth]{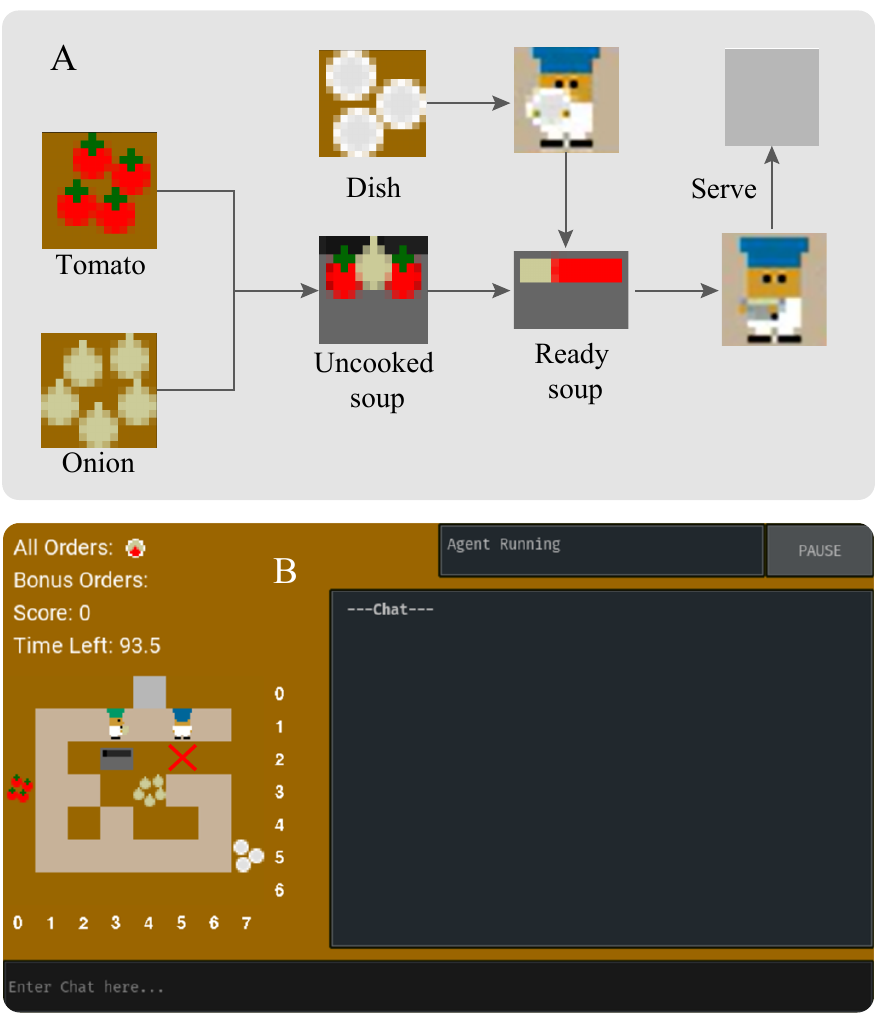}
\caption{(A) Cooking process to complete an order. (B) The designed human-AI collaboration interface (left: game layout, right: communication panel). The Blue hat agent and Green hat agent are operated by humans and robots, respectively. The red cross represents an example of the intermediate empty counter used to collaborate.}
\label{Fig. environment}
\end{figure}

To test the influence of language feedback on human-robot collaboration, we chose Overcooked-AI~\citep{carroll2020utility}, an environment designed to assess multi-agent coordination skills. In this game (\figref{Fig. environment}, humans (blue hat agents) and robots (green hat agents) are motivated to collaborate actively to maximize their score by completing orders within a time limit. A score of 53 points was awarded when the correct soup was served. Partial points will be awarded to incomplete or incorrect soups based on the number of missing/incorrect components. 
To cook a soup, chefs need to finish specific subtasks in sequence according to the recipes (see Fig. \ref{Fig. environment}A), and to finish subtasks, chefs need to move and interact with the environment. This decision making process can be applied to any collaborative setting: first, reasoning through subtasks to achieve the overall goal, then selecting low-level atomic actions to complete each subtask. 

\subsubsection{Subtasks}
The subtasks in an overcooked environment can be broken down into three main parts: 
\begin{enumerate}
    \item \textbf{Gathering Ingredients}: Chefs must first pick up the correct ingredients, such as onions or tomatoes, and place them into the cooking pot according to the recipe requirements.
    
    \item \textbf{Cooking}: Once the ingredients are placed in the pot, the chef has to start cooking. A timer on the pot signals when the soup is ready to be served.
    
    \item \textbf{Serving}: When the soup is ready, the chefs must collect and clean the dish from the dish dispenser, pour the soup into the dish, and deliver it to the serving location.
\end{enumerate}

In collaborative scenarios, the human and robot have a different path cost for completing each subtask, such as picking up an onion versus a tomato. Some subtasks may be unachievable for humans or robots. Furthermore, by decomposing a subtask into multiple smaller subtasks 
that can be performed by human and robot together, the overall time cost can be reduced. For example, in Fig. \ref{Fig. environment}, the robot might place the onion at the red cross point (5,2), allowing human to pick it up from there and add it to the pot.
\subsubsection{Atomic Action}
To finish a specific subtask, such as picking up an onion, the chef must execute a sequence of atomic actions, including movement commands like \textit{up}, \textit{down}, \textit{left}, \textit{right}, \textit{stay}, and \textit{interact} for picking up or placing objects.

\begin{figure*}[ht] 
\centering \vspace{0.08in} 
\includegraphics[width=0.98\linewidth]{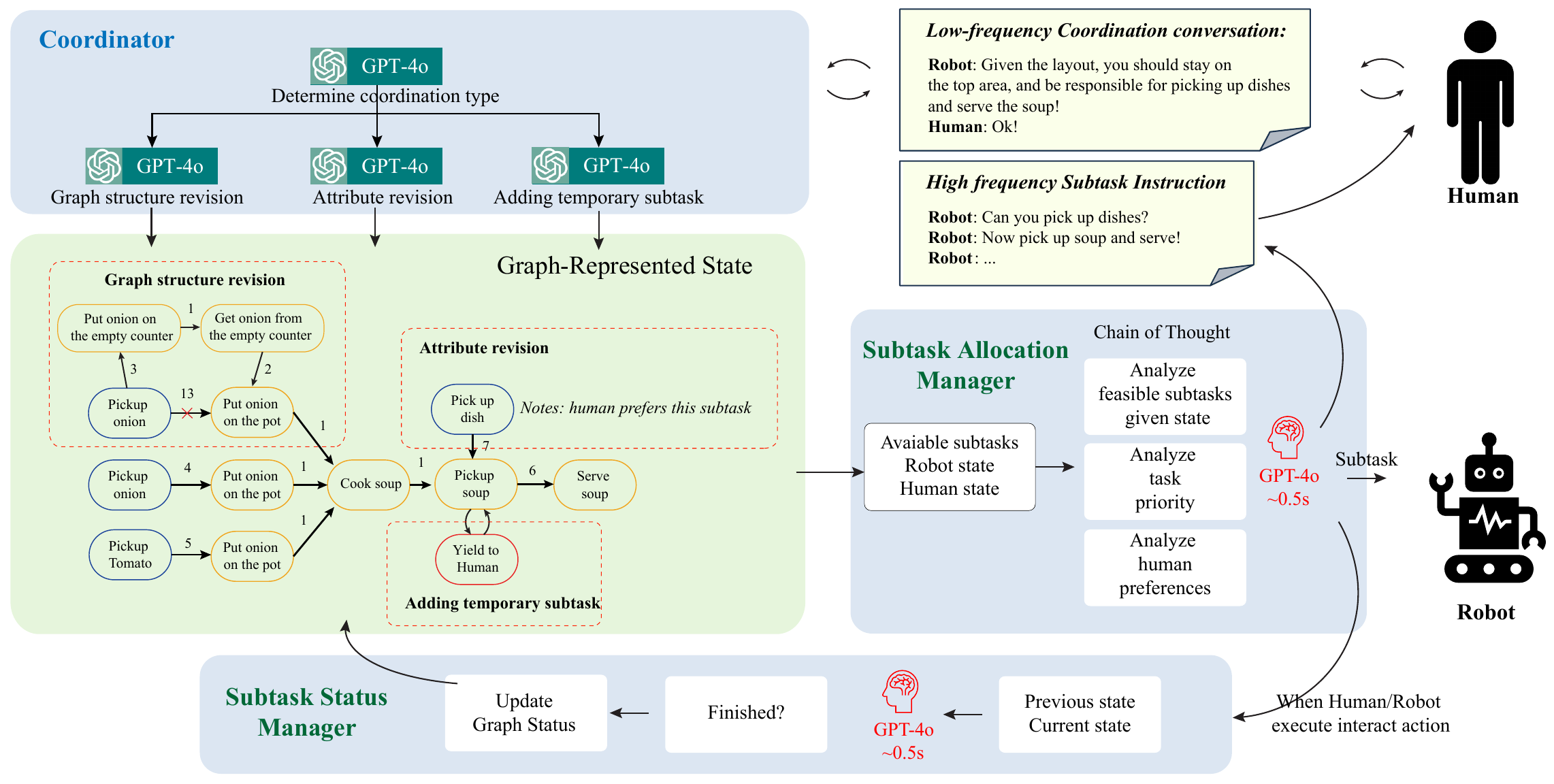} 
\caption{Human-Robot Teaming Framework with Multi-Modal Feedback(HRT-ML). It contains two modules: the \textit{Coordinator} and the \textit{Manager}. The \textit{Coordinator} designs overall coordination strategies using Directed Acyclic Graph (DAG) and dynamically add and delete nodes and edges through communication with humans. Three types of coordination—graph structure revision, graph node attribute revision, and adding temporary subtasks—are enclosed in a red dashed rectangle. A blue subtask node represents tasks that are ready for execution, a yellow node indicates tasks that are not yet ready, and a red node signifies emergency tasks. The action cost between two subtasks is indicated with numbers on the edges. The \textit{Manager} interprets the graph-represented coordination strategy to allocate subtasks to humans and robots. It provides subtask-level instructions to guide humans toward the overall goal and updates the graph status upon the completion of each subtask.} 
\label{fig.framework} 
\end{figure*}
\section{Human-Robot Teaming Framework with Multi-Modal Language Feedback}
To provide adaptive language feedback in human-robot collaboration, HRT-ML leverages a Directed Acyclic Graph (DAG) to represent subtasks and its dependencies (e.g., finishing an onion soup) for human-robot collaboration. Based on the DAG,  HRT-ML includes two core modules: the \textit{Coordinator} and the \textit{Manager}.  The \textit{Coordinator} handles high-level strategy discussions, generating an initial subtask graph and revising it based on conversation. The \textit{Manager} is responsible for allocating detailed subtasks based on the finalized coordination DAG, monitoring the status of assigned subtasks for both the human and the robot, and updating the graph accordingly. Once a subtask is determined, a \textit{Greedy Action Planner} is used to generate atomic actions for the robot.
\subsection{Subtask Graph for Human-Robot Collaboration}

We represent the collaboration goal as a dynamic directed acyclic graph (DAG) (\figref{fig.framework}) where nodes represent subtasks (e.g., \textit{pick onion}, \textit{cook onion}), and edges represent dependencies between tasks. Each edge is defined as \textit{(parent node,  node, cost)}, where the cost is calculated using the Greedy Action Planner (\secref{sec: planner}). Each node in the graph contains attributes for different coordination purposes:
\begin{itemize}
\item \textbf{Task id:} Each task gets a unique id, such that the LLM only needs to return the task id, and we can retrieve all other information based on the task id. 
\item \textbf{Task Type:} Categorized as \textit{Putting}, \textit{Getting}, or \textit{Operating}. There exist constraints when assigning different subtasks to robots or humans depending on their state, e.g., a robot that already held an onion can not execute any \textit{Getting} or \textit{Operating} tasks. 
\item \textbf{Status:} Represents the execution state, includes \textit{ready-to-execute}, \textit{executing}, \textit{success}, \textit{failure}, \textit{emergency}. The status of each subtask is considered when an LLM is prompted to assign subtasks for Humans and Robots. 
\item \textbf{Target Position:} Specifies a list of locations in the environment where the task can be performed. One subtask can be executed at multiple locations. This will enable humans and robots to explore additional collaboration strategies, such as the human placing an onion on an empty counter while the robot retrieves it.
\item \textbf{Notes:} The subtask node includes a notes attribute that captures human preferences or additional instructions (e.g., the human chef prefers to perform picking up an onion; the human chef is observed to be proficient at picking up onion.).
\item \textbf{Parent subtasks:} The parent subtasks indicate the prerequisite subtasks that are required to be \textit{success} before this subtask can be executed.
\item \textbf{Priority}: Priority is considered when assigning subtasks to humans and robots. It is calculated as the accumulated cost from the current node to the sink (termination) node. 
\item \textbf{Running time}: The running time acts as a counter, recording the duration for which the robot has been assigned to this subtask. 
\end{itemize}
The subtask graph enables the robot to translate language-based coordination into modifications within the graph, helping the LLM organize conversations and align its knowledge with the human's.
\subsection{Coordinator} 
The \textit{Coordinator} of the robot is responsible for designing overall collaboration strategies and discussing them with the human partner. The initial subtask graph is generated using a one-shot LLM prompt (\appref{app:prompt-gen}), incorporating recipe and kitchen layout information. We then implement a conversation-based graph modification, which can be initiated by the robot at a low frequency with the corresponding prompt (\appref{app:prompt-gen})  (active mode / superactive mode described in \secref{sec:modes} or proactively by the human through language-based feedback). The coordination process can be divided into the following decision tree:
\begin{itemize}
    \item The robot first identifies the type of human request, which falls into one of three categories (\figref{fig.framework}, red dashed rectangle); if it is unsure, it will further ask a human to clarify.
    \begin{itemize}
        \item \textit{Subtask Graph Structure Change:} The human expresses thoughts on how to complete the overall task. This may involve adding nodes, further decomposing a subtask for human and robot execution, or correcting incorrect graph dependencies. For example, based on the environment (\figref{Fig. environment}B), one of the possible coordination strategies is that the robot can place the onion at location (5, 2), and the human (blue hat) should focus on picking up onions at (5, 2).

        \item \textit{Subtask Attribute Change:} The human specifies preferences for subtask assignments. For example, based on the environment (\figref{Fig. environment}B), a human says that he would like to handle picking up onions and dishes as well as putting them on the counter (5, 2) and (6, 2).

        \item \textit{Temporary Subtask Assignment:} The human is not designing an overall collaboration strategy but instead requests the robot to execute a specific temporary subtask. For example, if the robot is blocking the human's path while they attempt to put an onion in the pot, the human may instruct the robot to move to a specific position. 
    \end{itemize}
    \item After identifying the type of coordination, the subtask graph is updated using the prompts shown in Appendix \ref{app:prompt-gen} for each category.
\end{itemize}

During these coordination discussions, the game remains paused until the human concludes the conversation. 

\subsection{Manager} 
The \textit{Manager} (\figref{fig.framework}) is responsible for real-time subtask allocation based on the information from the subtask graph and subtask status monitoring. 

\subsubsection{Subtask allocation}
The \textit{Manager} were first asked to analyze the current state of humans and robots along with the list of subtasks that are ready to execute in the subtask graph. The prompt (\appref{app:prompt-manager}) explicitly instructs the robot to consider task priorities and types based on the current state, as well as human-reported preferences for each subtask. In addition to subtask assignments, the \textit{Manager} is also prompted to generate language instructions to guide the human to execute their assigned subtasks toward the final goal. This feedback is selectively presented to the human based on the feedback mode described in \secref{sec:modes}. 

\subsubsection{Subtask status update}
The \textit{Manager} will be triggered to evaluate the current executing task to update the status given the human and robot action trajectory when an \textit{Interact} action is executed (Prompt details can be found in \appref{app:prompt-manager}). Once a subtask is identified as \textit{success}, the subtask nodes without unresolved dependencies will be marked as \textit{ready-to-execute}, while others remain \textit{not-ready} until their prerequisites are completed.

\subsection{Greedy Action Planner} 
\label{sec: planner}
Once a subtask (target location) is assigned, the robot uses a greedy planner based on Depth First Search (DFS) to determine the sequence of atomic actions required to complete the task. The planner computes the optimal path with the lowest action cost, considering the current environment layout and chef states.

\subsection{Multi-Modal Feedbacks}
\label{sec:modes}
Building on the proposed HRT-ML framework, we introduce robotic agents that provide four different language feedbacks described as follows: Inactive Feedback, Passive Feedback, Active Feedback, and Superactive Feedback. 
\begin{itemize}
    \item \textit{Inactive Feedback Agent (IFA)}: The IFA collaborates with humans without language communication and coordination. Only the Manager generates target subtasks for the greedy planner.  
    
    \item \textit{Passive Feedback Agent (PFA)}: The PFA starts to provide passive feedback only when a human requests. The human player takes the role of the leader, while the robotic agent acts as the follower, passively responding to human requests. If there are no specific human commands, PFA will behave like an IFA.
    
    \item \textit{Active Feedback Agent (AFA)}: The AFA collaborates with humans as peers. In this mode, both humans and robots can reach out to give language feedback. For user study, we prompt the GPT-4o coordinator in low frequency ($\sim$ 20s) to analyze human conversation history and suggest coordination strategies.
    
    \item \textit{Superactive Feedback Agent (SFA)}: The SFA treats humans as novices, acting as supervisors by providing frequent, continuous guidance on every subtask.
\end{itemize}

\section{Data collection}
In this section, we aim to explore the robot's ability to provide language feedback to help improve human satisfaction and teaming efficiency. Based on the proposed HRT-ML, we consider four types of language feedback and three different layouts: easy, medium, and hard, with increasing task difficulty and map complexity (Fig.~\ref{fig:survey-process}). Human participants will play on each layout paired with a robotic agent with varying levels of language feedback. Further details will be provided in the following section.

\subsection{Layouts with Different Complexities}
\begin{figure}[ht]
\centering
\vspace{0.08in}
\includegraphics[width=0.98\linewidth]{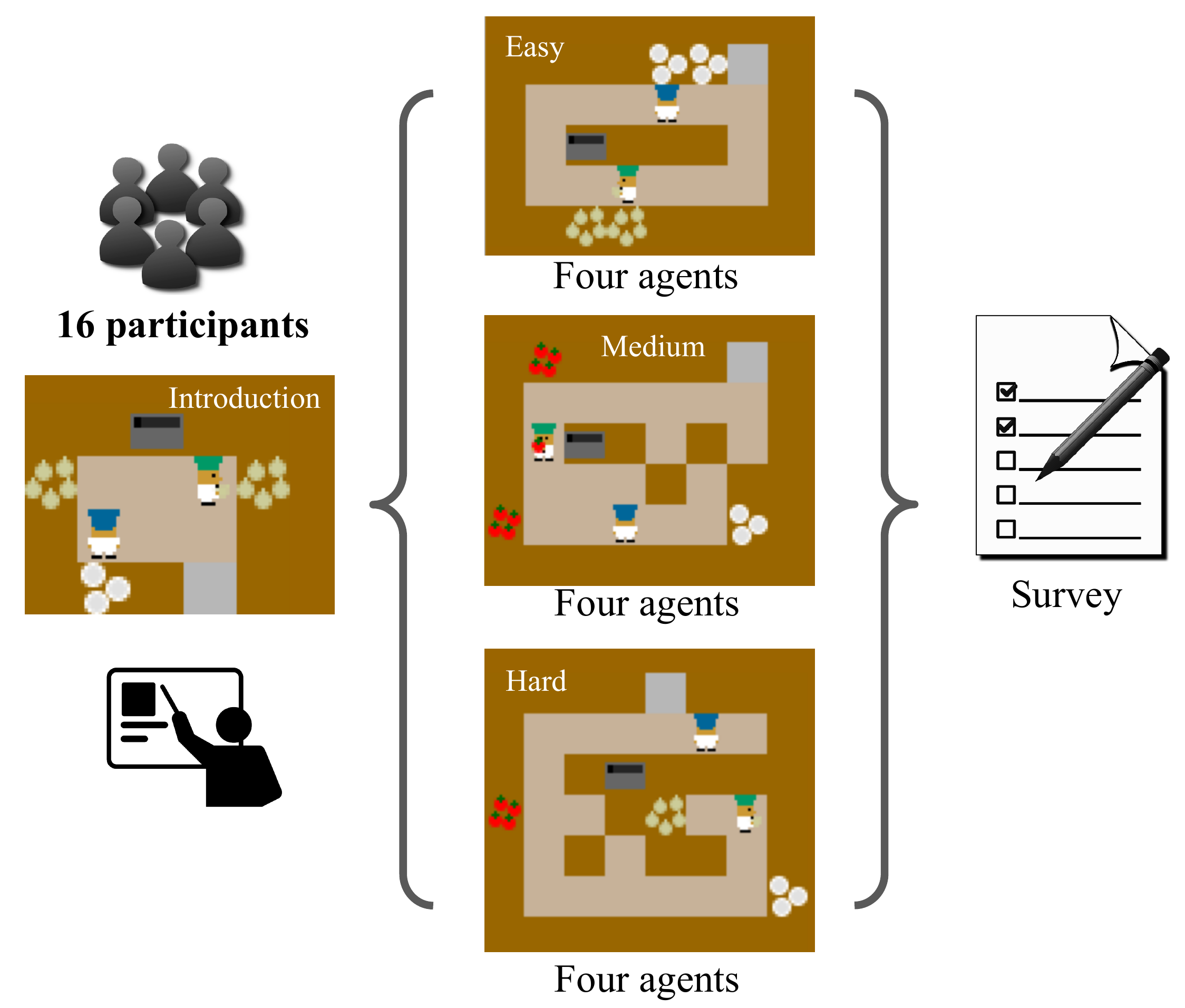}
\caption{Overview of the human study procedure involving 16 participants, the study begins with a tutorial map to introduce game mechanics. Followed by the tutorial, the human players will then play on the easy, medium, and hard layouts with four agents with different language feedback. Post-session surveys were conducted to collect data on participant satisfaction, engagement, trust, and feedback.}
\label{fig:survey-process}
\end{figure}

To test the performance of the four robotic agents and the influence of language feedback, we implemented four overcooked maps (Fig. \ref{fig:survey-process}). One of the maps is an introductory map, designed for participants to get familiar with the game and operation. The other three maps have varying levels of difficulty, which we refer to as easy, medium, and hard. The more challenging maps have more ``dead-ends'', requiring humans and robots to have a better coordination strategy to maintain team efficiency. Furthermore, the hard map introduces complexity in task orders by requiring multiple ingredients, which demands that humans and the robot reason about orders containing both tomatoes and onions. 

\subsection{Participants}
In this study, we recruited 16 participants (9 male and 7 female) aged between 23 and 33 to evaluate the performance of collaborative robotic agents. We selected participants with varying levels of familiarity with digital agents in-game environments: 10 participants self-reported video game time between 1-10 hours per week, 5 participants reported 10-20 hours per week, and 1 reported 20-30 hours per week. Additionally, 13 of the 16 participants reported prior familiarity with LLMs and embodied language agents, while 3 of the 16 participants reported no prior knowledge or experiences with language agents. 
\subsection{Procedures}

Participants were first asked to complete a consent form per the Institutional Review Board (IRB) protocol. 
Subsequently, prior to the formal trials they were given an opportunity to play with the four robotic agents in an introduction map (Fig.~\ref{fig:survey-process}) to explore various language feedback styles. During this introduction phase participants were guided through system operations, game rules, and robot functionalities. After the introduction phase, participants proceeded to independently collaborate with each of the four agent types on easy, medium, and hard layouts (Fig.~\ref{fig:survey-process}) to complete the maximum score within the given time limit (60 s for each layout). For each participant, the scenario (\textit{i.e.,} agent type $\times$ layout difficulty) was set to show up in a randomized order. We collected a total of 192 experiment trials, 12 trials per participant.

We collected game scores for each experiment shown in (Fig.~\ref{fig:survey-process}). After completing the teaming scenario, participants were also asked to fill out a survey rating their satisfaction, engagement, and trust level for each experiment, on a seven-point Likert scale. Participants were also asked to specify their preferred language feedback level for each layout.

Additionally, participants were asked to provide improvement suggestions on the language feedback provided by robots (\textit{e.g.,} \textit{``If you were to play with this robotic agent in an Overcooked game competition, what changes or improvements would you suggest for the agent's feedback?''}), and state the reason for their satisfaction ratings. For more details, the full questionnaire is available in the Appendix. \ref{sec:difficulty}.


\section{Results and Discussion}

The purpose of our data collection and analysis was to test the following hypotheses: 
\textit{An increased level of language support will result in an increase in the perceived level of trustworthiness and intelligence of the robot, and improve overall team performance.}

Surprisingly, our data suggested while language feedback could indeed facilitate human trust, perceived robot intelligence, and team efficiency, the desired level of language support exhibited a different relationship than hypothesized. We report our findings in the subsequent sections.

\subsection{Language Feedback Builds Human Trust and Perceived Intelligence}\label{sec:trust}
\begin{figure}[ht]
\centering
\vspace{0.08in}
\includegraphics[width=0.98\linewidth]{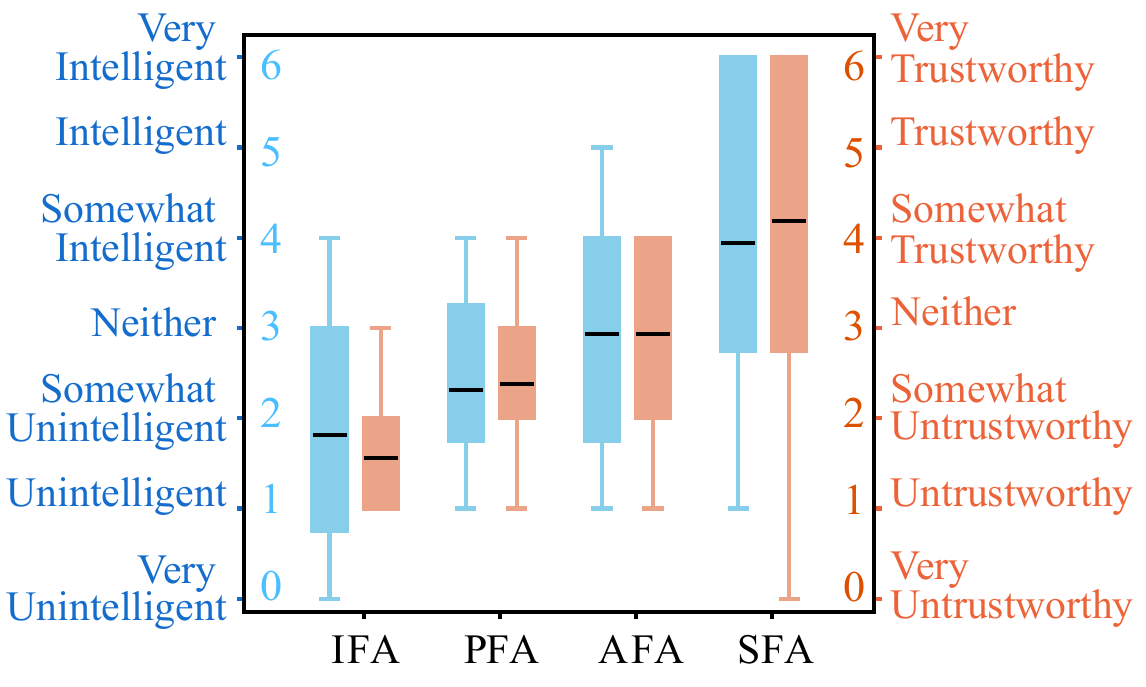}
\caption{Human perceived robot intelligence level (blue bar) and trust level (red bar) represented on a seven-point Likert scale, ranging from ``Very Unintelligent/Untrustworthy" to ``Very Intelligent/Trustworthy". }
\label{fig:trust-intelligence}
\end{figure}

Trust is a key factor in human-AI collaboration, shaping human experiences and significantly influencing long-term collaboration efficiency~\citep{chen2020trust}. In this section, we report the perceived levels of trust and intelligence after participants teamed with four types of robotic agents across three layouts. Both intelligence and trust levels were measured using a 7-point Likert scale, ranging from ``Very Untrustworthy'' (or ``Very Unintelligent'') to ``Very Trustworthy'' (or ``Very Intelligent''). We found that as the robot’s support level increased from Inactive to Superactive, intelligence rating increased monotonically (Fig. \ref{fig:trust-intelligence}). A similar trend was observed in the trust ratings. This suggested that as hypothesized, active communication can facilitate building trust and preceived intelligence. Another interesting observation was that, as the robot become more active, the standard deviation for trust ratings and intelligent ratings also increased. This suggested that both trust and intelligence levels also become more influenced by individual human preferences when the robot takes on a more active role in assisting. 

\subsection{Appropriate Language Feedback Improves Collaboration Efficiency}\label{sec:efficiency}
We used the game score to evaluate the team performance and collaboration efficiency between humans and robots. Overall, the team scored more points in the easy layout with all agent types (53.6 on average) as compared to the medium (30.6 on average) and hard (27.3 on average) layouts. This is not surprising, as the easy layout has simpler maps and fewer subtasks, requiring minimal coordination to complete. In contrast, medium and hard layouts introduced more complex subtasks and dependencies, requiring greater coordination and team effort, which could increase the chance of errors during the task and result in lower scores.

\begin{figure}[ht]
\centering
\vspace{0.08in}
\includegraphics[width=0.98\linewidth]{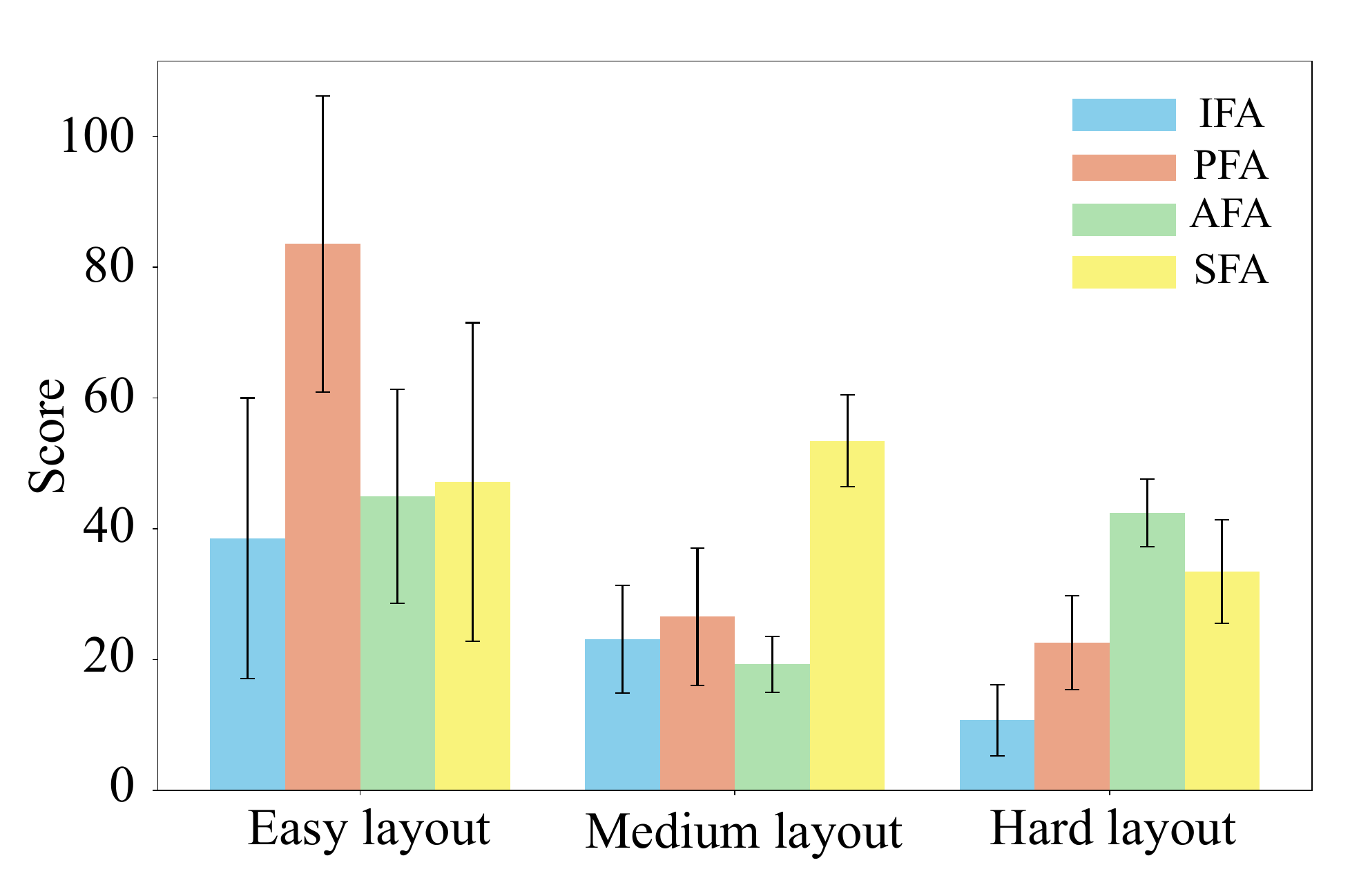}
\caption{
Game scores of all participants paired with different robotic agents across various layouts. The score distribution of each agent type, IFA, PFA, AFA, SFA, is represented by the blue, red, green, and yellow boxes, respectively.
}
\label{fig:score}
\end{figure}

We found that the team performance with ``active'' robotic agents (PFA, AFA, and SFA), which engaged in language feedback and coordination with human, achieved higher scores on almost all difficulty levels than the ``inactive'' agent (IFA), which did not engage in any communication (Fig. \ref{fig:score}). In addition, as the difficulty level increases from easy to hard, the team performance with the passive agent (PFA) decreased significantly, from close to 40 points to around 10 points (approximately 75$\%$ of performance drop). This result suggested that as hypothesized, language feedback can play a significant role in human-robot teaming, especially in complex tasks. 

However, the team performance was not always better with more active robotic agents. 
In the easy layout, team with PFA performed best, scoring 83.6 in average, significantly higher than IFA, AFA, and SFA (Fig. \ref{fig:score}). It was expected that PFA performed better than IFA in the easy task, as it provided effective support with minimal interference, enabling participants to benefit from its assistance as needed and allowing humans to take the lead. This was also supported by user feedback from the survey -- \textit{``Robot listening to the user and following the commands would help the game better. In terms of intelligence, passive robots work better but still lack user assistance"}. Unexpectedly, while AFA and SFA offered more active feedback, they achieved lower scores than the PFA (Fig. \ref{fig:score}), suggesting that constant communication was less effective for simple tasks, and may even distract humans. As the complexity increases to the medium level, the team with SFA exhibited a huge increase in score (Fig. \ref{fig:score}), exceeding the performance of PFA, implying that as task complexity increased, the frequent support provided by SFA became more valuable. Similarly, in the hard layout, the SFA and AFA demonstrated superior performance as compared to IFA and PFA (Fig. \ref{fig:score}), underscoring the value of active guidance and high-frequency support in facilitating collaboration and helping improve task execution in complex tasks.

Interestingly, despite SFA’s higher frequency of active support compared to AFA, overall performance still decreased due to the high complexity of the hard layout compared to AFA. Participant feedback highlighted this challenge, \textit{e.g.,} ``\textit{the robot/agent is not as smart as me,}" and, ``\textit{I have to give it a long instruction set, and I am more intelligent than it in the hard layout.}" One interpretation of these responses was that participants found the robot's support insufficient for complex tasks. As a result, instead of reducing cognitive load, the frequent suggestions from the robot required human to carefully think about responses, which ultimately increased cognitive demands and decreased team efficiency. Another interpretation is that psychologically, humans may have the preference to demonstrate and maintain intellectual superiority in challenging tasks when teaming with robots. As a result, how robotic agents communicate suggestions may greatly influence humans' acceptance rate and team efficiency.  

Overall, our results revealed that, the language feedback provided by LLMs can boost human-robot collaboration efficiency and increase human satisfaction. However, the proactiveness and frequency of the language feedback should be provided based on the task complexity and the capabilities of LLMs.

\subsection{Humans Don't Always Prefer the Best-performing Robotic Agents}\label{sec:SFA-performance}

\begin{figure}[ht]
\centering
\vspace{0.001in}
\includegraphics[width=1\linewidth]{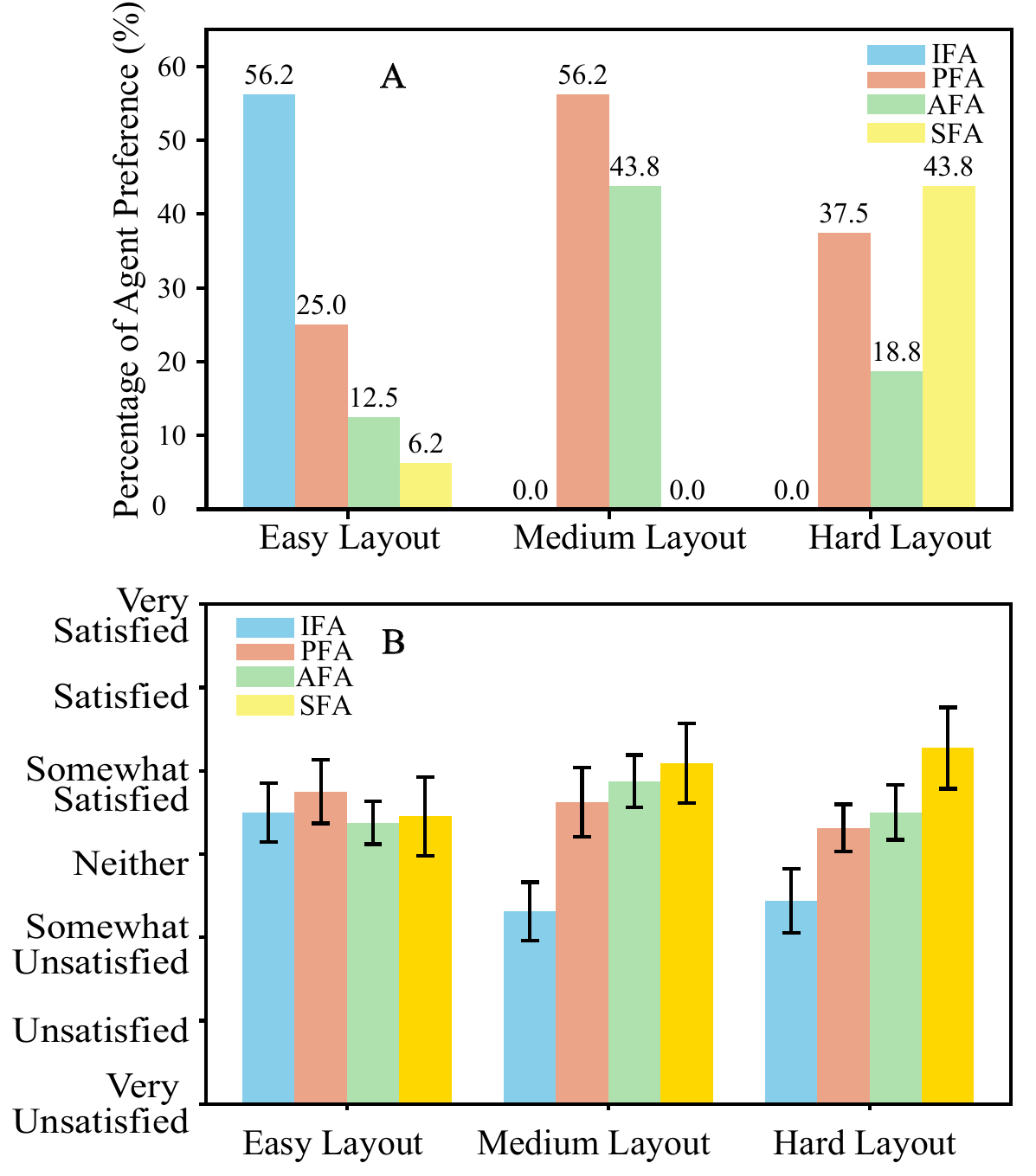}
\caption{Participant preferences and satisfaction levels for different agent types across layouts. (A) Bar chart showing the percentage of agent preference by participants for different layouts. (B) Satisfaction ratings for each agent type in different layouts, represented on a seven-point Likert scale, ranging from ``Very Unsatisfied" to ``Very Satisfied," with error bars indicating the standard error of the mean (SEM) in responses.}
\label{fig:overall-preference}
\end{figure}

Interestingly, our data suggested that human does not always prefer the agent type that helped achieve the highest game scores. For example, even though the IFA achieved the lowest scores on easy layout (Fig. \ref{fig:score}), over 50\% of participants reported IFA as their preferred agent among the four types (Fig. \ref{fig:overall-preference}A). Similarly, even though the SFA outperformed PFA by almost two folds in terms of team score (Fig. \ref{fig:score}), 56\% of participants selected the PFA as their preferred agent, and 0\% of them preferred SFA (Fig. \ref{fig:overall-preference}A). 

We believe this preference shift can be explained by the flow theory~\citep{csikszentmihalyi2000beyond,chen2024integrating}, which states that people feel most engaged when task complexity aligns with their skill level, and robots can help achieve this balance by adjusting their level of support. In the easy layout, the task was not significantly beyond human's skill level, and the need for additional help to maintain engagement was low. Therefore, while the active robotic agents can provide help, this help did not significantly influence human's engage level and satisfaction rate (Fig. \ref{fig:overall-preference}B). As a result, the non-active agent, which requires the lowest level of cognitive load (``cost'') was preferred. As the difficulty level increases, however, the gap between task complexity and human skill level increases, resulting a disruption to the task-capability balance and reduced engagement. Meanwhile, the additional feedback and support from the more active robotic agents, such as assigning subtasks (\textit{e.g.,} ``pick up the onion from (x, x).'', can reduce the coordination and planning effort on the human, and restore the task-capability balance and enhance human's feel of engagement and joy. At this point, the cost of communication became negligible as compared to the need for sense of achievement, making more active robotic agents more desirable in these challenging scenarios (Fig.~\ref{fig:overall-preference}B). 


\begin{figure}[ht]
\centering
\vspace{0.08in}
\includegraphics[width=0.98\linewidth]{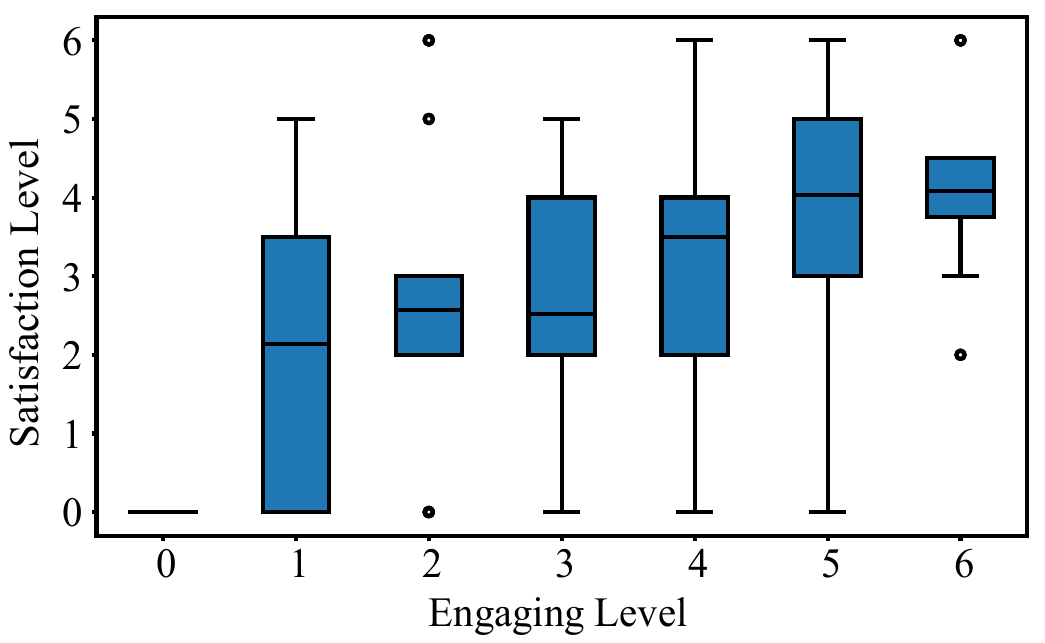}
\caption{Illustration of the relationship and variability between satisfaction level and the engagement of participants for all experimental trials. The box plot shows how satisfaction levels correspond to participants with an engagement level from 0 (Very distracted) to 6 (Very engaged).}
\label{fig:satvseng}
\end{figure}

To test this, we recorded participants' engagement during each game trial. As shown in Fig. \ref{fig:satvseng}, satisfaction levels increased with higher engagement, with a correlation test yielding a p-value of 0.00275 and a correlation coefficient of 0.93, indicating a strong relationship. This strong correlation between satisfaction and engagement indicates that engagement is the cause of different satisfaction levels reported by participants. This aligned with our theory, emphasizing the need to align robot support and feedback frequency based on task complexity relative to human capabilities.

\section{Adaptively assigning subtasks and providing language feedback}

Our findings revealed the importance of LLM-powered robots adapting their language feedback by considering the relationship between task complexity, $T_h$, human capability, $C_h$, and the LLM's capability, $C_l$. Below we propose a simple adaptation strategy for LLM-powered robots to select their support level and language feedback frequency: 

\begin{itemize}
    \item $C_h > T$ and  $C_l < T$: 
    Here human capability surpasses the task's complexity, and the LLM capability is not sufficient to address the challenging tasks without human guidance. Based on our results, a passive (PFA) to relative infrequent (AFA) feedback style would allow the robot to provide sufficient support to improve team performance and request human help when needed, while keeping communication frequency to a minimal to avoid overhead on the human side. This way, human teammates with higher capability level could guide the robot on high-level coordination strategies and specific subtask execution, keeping them engaged. 
    
    \item $C_h < T$ and $C_l > T$: 
    Here task complexity exceeds human capability, while LLM is fully capable of executing the task. In this case, extra active robot feedback and support (SFA) are crucial for maintaining team performance, and help reduce the gap between human capability and task complexity. 
    
    \item $C_h < T$ and $C_l < T$: 
    Here task complexity exceeds human capability. However, LLM capability is also not sufficient to address the challenging task neither. In this case, language feedback from LLM is often not useful in resolving the challenges that human is struggling with, and can even be misleading. High-frequency communications from LLM in this scenario would require additional effort from human to respond, potentially further heightening the anxiety that human is already experiencing~\citep{lenzner2010cognitive}, decreasing team performance and human engagement. Our results suggested that a more passive (PFA) or relative infrequent (AFA) feedback style would result in better teaming performance in this case.  

    \item $C_h > T$ and $C_l > T$: 
    Here both human and LLM capabilities surpass the task's complexity. The active feedback style (AFA) could allow the human and robots to communicate their needs at a comfortable pace, and improve collaboration efficiency. 
\end{itemize}


\section{Conclusion}

In this work, we introduced HRT-ML, a flexible human-robot teaming framework designed to provide adaptive communication feedback to humans at varying levels and frequencies. The HRT-ML framework comprises two core modules: a \textit{Coordinator} for high-level, low-frequency strategic guidance and a \textit{Manager} for task-specific, high-frequency instructions, allowing collaborating with humans across four distinct feedback styles: Inactive, Passive, Active, and Superactive.

Our user study results demonstrated that language-based feedback from LLMs can significantly enhance collaboration performance and foster human trust, and that as task complexity increases, more frequent, proactive support is desired. However, our study also revealed that it is critical for the robot to select their language feedback frequency based on task complexity, human capability, and robot capability. Overly frequent feedback in simple tasks or from less capable robots does not effectively increase team performance and satisfaction, and could even increase effort and reduce human engagement. Based on these findings, we proposed a simple principle that allows robots to adapt their language feedback style according to perceived task challenges, human capabilities, and LLM capabilities. 
 
\section{Limitations}
In this work, we focused on investigating the effect of robot feedback frequency on team performance. As such, robots were set to a constant active level and could not dynamically adjust their level of support and language feedback throughout the task. In real-world scenarios, task complexity and human skill levels for different sub-tasks can vary dynamically, requiring robots to adjust their communications and behaviors accordingly. 
Future work should explore methods for the adaptive robots to estimate human cognitive load, capabilities, and engagement and adjust LLM feedback in real-time to enable better team performance and adaptability. The results from our study provide the basis for designing and implementing such real-time feedback adjustments. Going forward, these adaptive response and feedback capabilities can empower future LLM-powered robots to flexibly support human needs in a wide variety of task complexities, fostering true partnerships and enhancing teamwork outcomes.

\section*{Acknowledgment}

We omitted the acknowledgment to comply with the double-blind review policy.

\bibliographystyle{plainnat}
\bibliography{main}

\begin{thebibliography}{39}
\providecommand{\natexlab}[1]{#1}
\providecommand{\url}[1]{\texttt{#1}}
\expandafter\ifx\csname urlstyle\endcsname\relax
  \providecommand{\doi}[1]{doi: #1}\else
  \providecommand{\doi}{doi: \begingroup \urlstyle{rm}\Url}\fi

\bibitem[Agashe et~al.(2023)Agashe, Fan, and Wang]{agashe2023evaluating}
Saaket Agashe, Yue Fan, and Xin~Eric Wang.
\newblock \href{https://arxiv.org/abs/2310.03903}{LLM-Coordination: Evaluating and Analyzing Multi-agent Coordination Abilities in Large Language Models}.
\newblock \emph{arXiv preprint arXiv:2310.03903}, 2023.

\bibitem[Ashktorab et~al.(2021)Ashktorab, Dugan, Johnson, Pan, Zhang, Kumaravel, and Campbell]{ashktorab2021effects}
Zahra Ashktorab, Casey Dugan, James Johnson, Qian Pan, Wei Zhang, Sadhana Kumaravel, and Murray Campbell.
\newblock \href{https://dl.acm.org/doi/10.1145/3411764.3445256}{Effects of communication directionality and AI agent differences in human-AI interaction}.
\newblock In \emph{Proceedings of the 2021 CHI conference on human factors in computing systems}, pages 1--15, 2021.

\bibitem[B{\i}y{\i}k et~al.(2022)B{\i}y{\i}k, Losey, Palan, Landolfi, Shevchuk, and Sadigh]{biyik2022learning}
Erdem B{\i}y{\i}k, Dylan~P Losey, Malayandi Palan, Nicholas~C Landolfi, Gleb Shevchuk, and Dorsa Sadigh.
\newblock \href{https://journals.sagepub.com/doi/10.1177/02783649211041652}{Learning reward functions from diverse sources of human feedback: Optimally integrating demonstrations and preferences}.
\newblock \emph{The International Journal of Robotics Research}, 41\penalty0 (1):\penalty0 45--67, 2022.

\bibitem[Bubeck et~al.(2023)Bubeck, Chandrasekaran, Eldan, Gehrke, Horvitz, Kamar, Lee, Lee, Li, Lundberg, et~al.]{bubeck2023sparks}
S{\'e}bastien Bubeck, Varun Chandrasekaran, Ronen Eldan, Johannes Gehrke, Eric Horvitz, Ece Kamar, Peter Lee, Yin~Tat Lee, Yuanzhi Li, Scott Lundberg, et~al.
\newblock \href{https://arxiv.org/abs/2303.12712}{Sparks of artificial general intelligence: Early experiments with gpt-4}.
\newblock \emph{arXiv preprint arXiv:2303.12712}, 2023.

\bibitem[Carroll et~al.(2019)Carroll, Shah, Ho, Griffiths, Seshia, Abbeel, and Dragan]{carroll2020utility}
Micah Carroll, Rohin Shah, Mark~K Ho, Tom Griffiths, Sanjit Seshia, Pieter Abbeel, and Anca Dragan.
\newblock \href{https://proceedings.neurips.cc/paper/2019/hash/f5b1b89d98b7286673128a5fb112cb9a-Abstract.html}{On the utility of learning about humans for human-ai coordination}.
\newblock \emph{Advances in neural information processing systems}, 32, 2019.

\bibitem[Chang et~al.(2024)Chang, Chhablani, Clegg, Cote, Desai, Hlavac, Karashchuk, Krantz, Mottaghi, Parashar, et~al.]{chang2024partnr}
Matthew Chang, Gunjan Chhablani, Alexander Clegg, Mikael~Dallaire Cote, Ruta Desai, Michal Hlavac, Vladimir Karashchuk, Jacob Krantz, Roozbeh Mottaghi, Priyam Parashar, et~al.
\newblock \href{https://arxiv.org/abs/2411.00081}{PARTNR: A Benchmark for Planning and Reasoning in Embodied Multi-agent Tasks}.
\newblock \emph{arXiv preprint arXiv:2411.00081}, 2024.

\bibitem[Chen et~al.(2024)Chen, Alghowinem, Breazeal, and Park]{chen2024integrating}
Huili Chen, Sharifa Alghowinem, Cynthia Breazeal, and Hae~Won Park.
\newblock \href{https://arxiv.org/abs/2401.02833}{Integrating Flow Theory and Adaptive Robot Roles: A Conceptual Model of Dynamic Robot Role Adaptation for the Enhanced Flow Experience in Long-term Multi-person Human-Robot Interactions}.
\newblock In \emph{Proceedings of the 2024 ACM/IEEE International Conference on Human-Robot Interaction}, pages 116--126, 2024.

\bibitem[Chen et~al.(2020)Chen, Nikolaidis, Soh, Hsu, and Srinivasa]{chen2020trust}
Min Chen, Stefanos Nikolaidis, Harold Soh, David Hsu, and Siddhartha Srinivasa.
\newblock \href{https://dl.acm.org/doi/10.1145/3359616}{Trust-aware decision making for human-robot collaboration: Model learning and planning}.
\newblock \emph{ACM Transactions on Human-Robot Interaction (THRI)}, 9\penalty0 (2):\penalty0 1--23, 2020.

\bibitem[Christoforou et~al.(2020)Christoforou, Panayides, Avgousti, Masouras, and Pattichis]{christoforou2020overview}
Eftychios~G Christoforou, Andreas~S Panayides, Sotiris Avgousti, Panicos Masouras, and Constantinos~S Pattichis.
\newblock \href{https://link.springer.com/chapter/10.1007/978-3-030-31635-8_118}{An overview of assistive robotics and technologies for elderly care}.
\newblock In \emph{XV Mediterranean Conference on Medical and Biological Engineering and Computing--MEDICON 2019: Proceedings of MEDICON 2019, September 26-28, 2019, Coimbra, Portugal}, pages 971--976. Springer, 2020.

\bibitem[Chuah and Yu(2021)]{chuah2021future}
Stephanie Hui-Wen Chuah and Joanne Yu.
\newblock \href{https://www.sciencedirect.com/science/article/pii/S096969892100117X}{The future of service: The power of emotion in human-robot interaction}.
\newblock \emph{Journal of Retailing and Consumer Services}, 61:\penalty0 102551, 2021.

\bibitem[Csikszentmihalyi(2000)]{csikszentmihalyi2000beyond}
Mihaly Csikszentmihalyi.
\newblock \emph{\href{https://archive.org/details/beyondboredomanx00csik}{Beyond boredom and anxiety.}}
\newblock Jossey-bass, 2000.

\bibitem[Deng et~al.(2023)Deng, Lei, Huang, and Chua]{deng2023rethinking}
Yang Deng, Wenqiang Lei, Minlie Huang, and Tat-Seng Chua.
\newblock \href{https://dl.acm.org/doi/10.1145/3624918.3629548}{Rethinking Conversational Agents in the Era of LLMs: Proactivity, Non-collaborativity, and Beyond}.
\newblock In \emph{Proceedings of the Annual International ACM SIGIR Conference on Research and Development in Information Retrieval in the Asia Pacific Region}, pages 298--301, 2023.

\bibitem[Faccio et~al.(2023)Faccio, Granata, Menini, Milanese, Rossato, Bottin, Minto, Pluchino, Gamberini, Boschetti, et~al.]{faccio2023human}
Maurizio Faccio, Irene Granata, Alberto Menini, Mattia Milanese, Chiara Rossato, Matteo Bottin, Riccardo Minto, Patrik Pluchino, Luciano Gamberini, Giovanni Boschetti, et~al.
\newblock \href{https://link.springer.com/article/10.1007/s10845-022-01953-w}{Human factors in cobot era: a review of modern production systems features}.
\newblock \emph{Journal of Intelligent Manufacturing}, 34\penalty0 (1):\penalty0 85--106, 2023.

\bibitem[Ge et~al.(2023)Ge, Hua, Mei, Tan, Xu, Li, Zhang, et~al.]{ge2024openagi}
Yingqiang Ge, Wenyue Hua, Kai Mei, Juntao Tan, Shuyuan Xu, Zelong Li, Yongfeng Zhang, et~al.
\newblock Openagi: When llm meets domain experts.
\newblock \emph{Advances in Neural Information Processing Systems}, 36:\penalty0 5539--5568, 2023.

\bibitem[Gordon et~al.(2020)Gordon, Meng, Bhattacharjee, Barnes, and Srinivasa]{gordon2020adaptive}
Ethan~K Gordon, Xiang Meng, Tapomayukh Bhattacharjee, Matt Barnes, and Siddhartha~S Srinivasa.
\newblock \href{https://ieeexplore.ieee.org/document/9341359}{Adaptive robot-assisted feeding: An online learning framework for acquiring previously unseen food items}.
\newblock In \emph{2020 IEEE/RSJ International Conference on Intelligent Robots and Systems (IROS)}, pages 9659--9666. IEEE, 2020.

\bibitem[Guan et~al.(2023)Guan, Zhang, Fan, Li, Chen, Li, Tian, Yuan, and Yu]{guan2023efficient}
Cong Guan, Lichao Zhang, Chunpeng Fan, Yichen Li, Feng Chen, Lihe Li, Yunjia Tian, Lei Yuan, and Yang Yu.
\newblock \href{https://arxiv.org/abs/2311.00416}{Efficient Human-AI Coordination via Preparatory Language-based Convention}.
\newblock \emph{arXiv preprint arXiv:2311.00416}, 2023.

\bibitem[Hoffman(2019)]{hoffman2019evaluating}
Guy Hoffman.
\newblock \href{https://ieeexplore.ieee.org/document/8678448}{Evaluating fluency in human--robot collaboration}.
\newblock \emph{IEEE Transactions on Human-Machine Systems}, 49\penalty0 (3):\penalty0 209--218, 2019.

\bibitem[Hong et~al.(2023)Hong, Zheng, Chen, Cheng, Wang, Zhang, Wang, Yau, Lin, Zhou, et~al.]{hong2023metagpt}
Sirui Hong, Xiawu Zheng, Jonathan Chen, Yuheng Cheng, Jinlin Wang, Ceyao Zhang, Zili Wang, Steven Ka~Shing Yau, Zijuan Lin, Liyang Zhou, et~al.
\newblock \href{https://arxiv.org/abs/2308.00352}{Metagpt: Meta programming for multi-agent collaborative framework}.
\newblock \emph{arXiv preprint arXiv:2308.00352}, 2023.

\bibitem[Hou et~al.(2024)Hou, Tamoto, and Miyashita]{hou2024my}
Yuki Hou, Haruki Tamoto, and Homei Miyashita.
\newblock \href{https://dl.acm.org/doi/10.1145/3613905.3650839}{" My agent understands me better": Integrating Dynamic Human-like Memory Recall and Consolidation in LLM-Based Agents}.
\newblock In \emph{Extended Abstracts of the CHI Conference on Human Factors in Computing Systems}, pages 1--7, 2024.

\bibitem[Jahanmahin et~al.(2022)Jahanmahin, Masoud, Rickli, and Djuric]{jahanmahin2022human}
Roohollah Jahanmahin, Sara Masoud, Jeremy Rickli, and Ana Djuric.
\newblock \href{https://www.sciencedirect.com/science/article/abs/pii/S0736584522000916?via%3Dihub}{Human-robot interactions in manufacturing: A survey of human behavior modeling}.
\newblock \emph{Robotics and Computer-Integrated Manufacturing}, 78:\penalty0 102404, 2022.

\bibitem[Lenzner et~al.(2010)Lenzner, Kaczmirek, and Lenzner]{lenzner2010cognitive}
Timo Lenzner, Lars Kaczmirek, and Alwine Lenzner.
\newblock \href{https://doi.org/10.1002/acp.1602}{Cognitive burden of survey questions and response times: A psycholinguistic experiment}.
\newblock \emph{Applied cognitive psychology}, 24\penalty0 (7):\penalty0 1003--1020, 2010.

\bibitem[Li et~al.(2024)Li, Sun, Chen, Fan, Wang, Liu, Zhu, Gan, and Tan]{li2024conav}
Changhao Li, Xinyu Sun, Peihao Chen, Jugang Fan, Zixu Wang, Yanxia Liu, Jinhui Zhu, Chuang Gan, and Mingkui Tan.
\newblock \href{https://arxiv.org/abs/2406.02425}{CoNav: A Benchmark for Human-Centered Collaborative Navigation}.
\newblock \emph{arXiv preprint arXiv:2406.02425}, 2024.

\bibitem[Liu et~al.(2023{\natexlab{a}})Liu, Yu, Gao, Xie, Liao, Wu, and Wang]{liu2023llm}
Jijia Liu, Chao Yu, Jiaxuan Gao, Yuqing Xie, Qingmin Liao, Yi~Wu, and Yu~Wang.
\newblock \href{https://www.ifaamas.org/Proceedings/aamas2024/pdfs/p1219.pdf}{Llm-powered hierarchical language agent for real-time human-ai coordination}.
\newblock \emph{arXiv preprint arXiv:2312.15224}, 2023{\natexlab{a}}.

\bibitem[Liu et~al.(2023{\natexlab{b}})Liu, Wilson, Krishnamachari, and Qian]{Liu2023}
Shipeng Liu, Cristina~G Wilson, Bhaskar Krishnamachari, and Feifei Qian.
\newblock \href{https://dl.acm.org/doi/10.1145/3623383}{Understanding Human Dynamic Sampling Objectives to Enable Robot-assisted Scientific Decision Making}.
\newblock \emph{ACM Transactions on Human-Robot Interaction}, 2023{\natexlab{b}}.

\bibitem[Liu et~al.(2024)Liu, Wilson, Lee, and Qian]{10.1145/3610977.3635112}
Shipeng Liu, Cristina~G. Wilson, Zachary~I. Lee, and Feifei Qian.
\newblock \href{https://dl.acm.org/doi/10.1145/3610977.3635112}{Modelling Experts' Sampling Strategy to Balance Multiple Objectives During Scientific Explorations}.
\newblock In \emph{Proceedings of the 2024 ACM/IEEE International Conference on Human-Robot Interaction}, HRI '24, page 452–461, New York, NY, USA, 2024. Association for Computing Machinery.
\newblock ISBN 9798400703225.
\newblock \doi{10.1145/3610977.3635112}.
\newblock URL \url{https://doi.org/10.1145/3610977.3635112}.

\bibitem[Lu et~al.(2024)Lu, Yang, Qian, Chen, Luo, Wu, Wang, Cong, Zhang, Lin, et~al.]{lu2024proactive}
Yaxi Lu, Shenzhi Yang, Cheng Qian, Guirong Chen, Qinyu Luo, Yesai Wu, Huadong Wang, Xin Cong, Zhong Zhang, Yankai Lin, et~al.
\newblock \href{https://arxiv.org/abs/2410.12361}{Proactive Agent: Shifting LLM Agents from Reactive Responses to Active Assistance}.
\newblock \emph{arXiv preprint arXiv:2410.12361}, 2024.

\bibitem[Ouyang et~al.(2022)Ouyang, Wu, Jiang, Almeida, Wainwright, Mishkin, Zhang, Agarwal, Slama, Ray, et~al.]{ouyang2022training}
Long Ouyang, Jeffrey Wu, Xu~Jiang, Diogo Almeida, Carroll Wainwright, Pamela Mishkin, Chong Zhang, Sandhini Agarwal, Katarina Slama, Alex Ray, et~al.
\newblock \href{https://proceedings.neurips.cc/paper_files/paper/2022/hash/b1efde53be364a73914f58805a001731-Abstract-Conference.html}{Training language models to follow instructions with human feedback}.
\newblock \emph{Advances in neural information processing systems}, 35:\penalty0 27730--27744, 2022.

\bibitem[{\"O}zdemir et~al.(2022){\"O}zdemir, Kerzel, Weber, Lee, and Wermter]{ozdemir2022language}
Ozan {\"O}zdemir, Matthias Kerzel, Cornelius Weber, Jae~Hee Lee, and Stefan Wermter.
\newblock \href{https://ieeexplore.ieee.org/document/9515668}{Language-Model-Based Paired Variational Autoencoders for Robotic Language Learning}.
\newblock \emph{IEEE Transactions on Cognitive and Developmental Systems}, 15\penalty0 (4):\penalty0 1812--1824, 2022.

\bibitem[Park et~al.(2020)Park, Hoshi, Mahajan, Kim, Erickson, Rogers, and Kemp]{park2020active}
Daehyung Park, Yuuna Hoshi, Harshal~P Mahajan, Ho~Keun Kim, Zackory Erickson, Wendy~A Rogers, and Charles~C Kemp.
\newblock Active robot-assisted feeding with a general-purpose mobile manipulator: Design, evaluation, and lessons learned.
\newblock \emph{Robotics and Autonomous Systems}, 124:\penalty0 103344, 2020.

\bibitem[Schoenegger et~al.(2024)Schoenegger, Park, Karger, Trott, and Tetlock]{schoenegger2024ai}
Philipp Schoenegger, Peter~S Park, Ezra Karger, Sean Trott, and Philip~E Tetlock.
\newblock \href{https://dl.acm.org/doi/abs/10.1145/3707649}{Ai-augmented predictions: Llm assistants improve human forecasting accuracy}.
\newblock \emph{arXiv preprint arXiv:2402.07862}, 2024.

\bibitem[Sharma et~al.(2022)Sharma, Sundaralingam, Blukis, Paxton, Hermans, Torralba, Andreas, and Fox]{sharma2022correcting}
Pratyusha Sharma, Balakumar Sundaralingam, Valts Blukis, Chris Paxton, Tucker Hermans, Antonio Torralba, Jacob Andreas, and Dieter Fox.
\newblock \href{https://www.roboticsproceedings.org/rss18/p065.pdf}{Correcting robot plans with natural language feedback}.
\newblock \emph{arXiv preprint arXiv:2204.05186}, 2022.

\bibitem[Sun et~al.(2024)Sun, Salami~Pargoo, Jin, and Ortiz]{sun2024optimizing}
Yuan Sun, Navid Salami~Pargoo, Peter Jin, and Jorge Ortiz.
\newblock Optimizing autonomous driving for safety: A human-centric approach with llm-enhanced rlhf.
\newblock In \emph{Companion of the 2024 on ACM International Joint Conference on Pervasive and Ubiquitous Computing}, pages 76--80, 2024.

\bibitem[Tanneberg et~al.(2024)Tanneberg, Ocker, Hasler, Deigmoeller, Belardinelli, Wang, Wersing, Sendhoff, and Gienger]{tanneberg2024help}
Daniel Tanneberg, Felix Ocker, Stephan Hasler, Joerg Deigmoeller, Anna Belardinelli, Chao Wang, Heiko Wersing, Bernhard Sendhoff, and Michael Gienger.
\newblock \href{https://arxiv.org/abs/2403.12533}{To Help or Not to Help: LLM-based Attentive Support for Human-Robot Group Interactions}.
\newblock \emph{arXiv preprint arXiv:2403.12533}, 2024.

\bibitem[Wu et~al.(2023)Wu, Bansal, Zhang, Wu, Zhang, Zhu, Li, Jiang, Zhang, and Wang]{wuautogen}
Qingyun Wu, Gagan Bansal, Jieyu Zhang, Yiran Wu, Shaokun Zhang, Erkang Zhu, Beibin Li, Li~Jiang, Xiaoyun Zhang, and Chi Wang.
\newblock \href{https://arxiv.org/pdf/2308.08155}{Autogen: Enabling next-gen llm applications via multi-agent conversation framework}.
\newblock \emph{arXiv preprint arXiv:2308.08155}, 2023.

\bibitem[Xiao et~al.(2020)Xiao, Wang, Lu, and Zhang]{xiao2020three}
Junhao Xiao, Pan Wang, Huimin Lu, and Hui Zhang.
\newblock \href{https://journals.sagepub.com/doi/full/10.1177/1729881420925293}{A three-dimensional mapping and virtual reality-based human--robot interaction for collaborative space exploration}.
\newblock \emph{International Journal of Advanced Robotic Systems}, 17\penalty0 (3):\penalty0 1729881420925293, 2020.

\bibitem[Zhang et~al.(2023{\natexlab{a}})Zhang, Yang, Hu, Wang, Li, Sun, Zhang, Zhang, Liu, Zhu, et~al.]{zhang2023proagent}
Ceyao Zhang, Kaijie Yang, Siyi Hu, Zihao Wang, Guanghe Li, Yihang Sun, Cheng Zhang, Zhaowei Zhang, Anji Liu, Song-Chun Zhu, et~al.
\newblock \href{https://dl.acm.org/doi/10.1609/aaai.v38i16.29710}{Proagent: Building proactive cooperative ai with large language models}.
\newblock \emph{arXiv preprint arXiv:2308.11339}, 2023{\natexlab{a}}.

\bibitem[Zhang et~al.(2023{\natexlab{b}})Zhang, Du, Shan, Zhou, Du, Tenenbaum, Shu, and Gan]{zhang2023building}
Hongxin Zhang, Weihua Du, Jiaming Shan, Qinhong Zhou, Yilun Du, Joshua~B Tenenbaum, Tianmin Shu, and Chuang Gan.
\newblock \href{https://arxiv.org/pdf/2307.02485}{Building cooperative embodied agents modularly with large language models}.
\newblock \emph{arXiv preprint arXiv:2307.02485}, 2023{\natexlab{b}}.

\bibitem[Zhang et~al.(2024{\natexlab{a}})Zhang, Wang, Lyu, Zhang, Chen, Shu, Du, and Gan]{zhang2024combo}
Hongxin Zhang, Zeyuan Wang, Qiushi Lyu, Zheyuan Zhang, Sunli Chen, Tianmin Shu, Yilun Du, and Chuang Gan.
\newblock \href{https://arxiv.org/pdf/2404.10775}{COMBO: Compositional World Models for Embodied Multi-Agent Cooperation}.
\newblock \emph{arXiv preprint arXiv:2404.10775}, 2024{\natexlab{a}}.

\bibitem[Zhang et~al.(2024{\natexlab{b}})Zhang, Wang, Zhang, Chen, Gao, Wang, Zhang, Wang, and Wen]{zhang2024mutual}
Shao Zhang, Xihuai Wang, Wenhao Zhang, Yongshan Chen, Landi Gao, Dakuo Wang, Weinan Zhang, Xinbing Wang, and Ying Wen.
\newblock \href{https://arxiv.org/pdf/2409.08811}{Mutual theory of mind in human-ai collaboration: An empirical study with llm-driven ai agents in a real-time shared workspace task}.
\newblock \emph{arXiv preprint arXiv:2409.08811}, 2024{\natexlab{b}}.

\end{thebibliography}



\clearpage

\appendices
\section{Overcooked Environment Layout Complexity} 
\label{sec:layout}
\begin{figure}[ht]
\centering
\vspace{0.08in}
\includegraphics[width=0.98\linewidth]{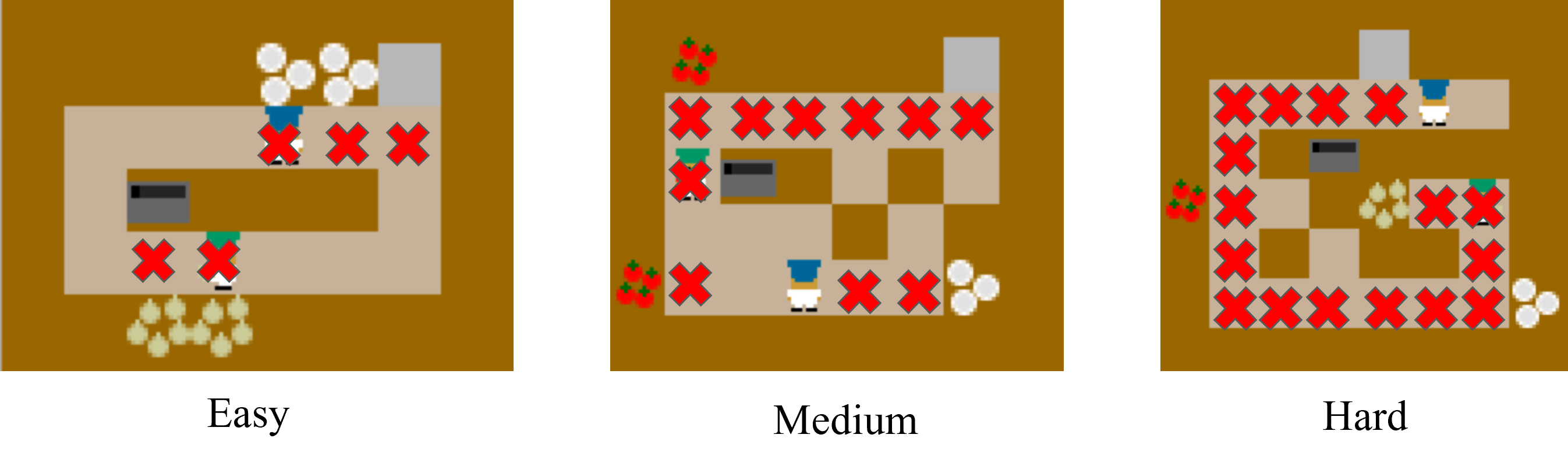}
\caption{Easy, medium, hard layouts. The red crosses represent the critical points that would prevent others from accomplishing any of the subtasks.}
\label{fig:complexity}
\end{figure}

To design layouts of varying difficulty in Overcooked game, we adopt teaming fluency metrics discussed by Hoffman~\cite{hoffman2019evaluating}. We calculate layout teaming fluency as the percentage of unobstructed areas within a layout's total free area. An obstructed area is defined as a critical position that would prevent others from completing their current subtasks. We hand-designed 22 layouts based on the teaming fluency and selected three representative layouts to conduct the user study. The teaming fluency metrics for the easy, medium, and hard maps are 64.29\%, 44.44\%, 20\%, respectively. As shown in~\figref{fig:complexity}, red crosses represent critical points of a layout. 

\section{User study Survey} 
\label{sec:difficulty}
We designed a questionnaire and collected user responses after playing the game. The survey collected data including user agreement, demographic information, satisfaction level, engagement level, agent preference, intelligent level, and trust level, as well as some open-ended questions. The following shows detailed questions in the questionnaire.
\begin{enumerate}
    \item \textbf{User on-boarding (goals, rules, consent)}
     \item \textbf{Participant demographics}
    \begin{enumerate}
        \item What is your age?
        \item What is your gender?
        \item How often do you play video games every week?
        \item Are you familiar with the usage of large language models?  embodied agents? 
    \end{enumerate}

    \item \textbf{Satisfaction and engagement}
    \begin{enumerate}
        \item Describe the satisfactory level of different agents in different layouts from 1 (Very unsatisfied) to 7 (very satisfied).
        \item Please describe the reason for the different level of satisfaction
        \item Describe the engaging level of you when you interacts with different agents in different layouts from 1 (very distracted) to 7 (very engaged).

    \end{enumerate}
    \item \textbf{Preference}
    \begin{enumerate}
        \item In layout "easy", which agent do you prefer? 
        \item In layout "medium", which agent do you prefer? 
        \item In layout "hard", which agent do you prefer? 
    \end{enumerate}

    \item \textbf{Perceived intelligence}
    \begin{enumerate}
        \item Please evaluate the intelligence level of three different agents from 1 (very unintelligent) to 7 (very intelligent).
    \end{enumerate}
    \item \textbf{Perceived trust}
    \begin{enumerate}
        \item  Please indicate the level of trust you have in three different agents to work with you in the Overcooked game from 1 (very trustworthy) to 7 (very untrustworthy).
    \end{enumerate}
    \item \textbf{Miscellaneous}
    \begin{enumerate}
        \item If you were to play with this agent in an Overcooked game competition, what changes or improvements would you suggest for the agent's feedback?
    \end{enumerate}
\end{enumerate}

\section{Prompt Construction} 
We include the full content of the prompts used by the \textit{Coordinator} and the \textit{Manager}. In particular, 
\begin{enumerate}
    \item \textit{Coordinator} generates the initial node graph and revises it with a human player through natural language dialog. we show prompts for initial graph generation in~\appref{app:prompt-gen}. we also show examples of revising the node graph by adding intermediate nodes, human preferences through natural language dialog, and how they actively suggest new coordination strategies in~\appref{app:prompt-gen}.
    \item \textit{Manager} assigns subtasks to players and changes player status. We show the prompts used for Manager to actively assign subtasks and change subtask status in~\appref{app:prompt-manager}.
\end{enumerate}
 
\subsection{Coordinaotor: Node Graph Generation and Revision}\label{app:prompt-gen}

This section provides the prompts for all possible nodes to exist in the behavior tree. Note that to add a new behavior, one simply needs to create a prompt for the behavior and add it as an option for other relevant behaviors to invoke.

\textbf{Initial Graph Generation}
The goal of this prompt is to generate an initial node graph for revision and execution
\begin{lstlisting}
Instructions: |
    The normal procedure to finish one soup is:
    1. Pick up the required ingredients 
    2. Put ingredients into pots.
    3. start cook the soup.
    4. Pick up a dish.
    5. pick up the soup after it is ready.
    6. Put the soup to the serve location.
    
    Remember, you must put the all ingredients and correct ingredients exactly as specified in the recipe book, a important thing that you have to put the soup to the serve location
    Recipe book:
    {recipe_book}
    
    Kitchen state:
    {kitchen_items}
        
    Available_subtask_types:
    0: "PUTTING", 1: "GETTING", 2: "COOKING"
    
    Available_subtask_status:
    0: "UNKNOWN", 1: "READY_TO_EXECUTE", 2: "SUCCESS", 3: "FAIL", 4: "NOT READY", 5"EXECUTING"
    
    Example subtasks output form:
    example_subtask = {
        "id": int,  # Unique ID of the subtask start from 0
        "name": string,  # Task description, e.g. "Get onion", the task can onlyh be three forms, pick, put, start_to_cook
        "target_position_id": list[int],  # IDs of target positions selected from provided locations
        "task_type": int,  # Integer representing the task type (e.g., 1 = GETTING, refer to all avaiable types)
        "task_status": int,  # Integer representing the task status (e.g., refer to all avaiable status, but you only to judge if this subtask has been finished, if not, leave unknown, I will handle it based on graph)
        "notes": str, # if human has some preferences related to this subtask, you should write it in a very short sentences here, e.g. human preferes to do this task
        "parent_subtask": list[int]  #  Only list of IDs of parent subtasks that are a must and reprequisite to this task, (leave empty if no required subtasks before this, or other agent can help do this)
            }
    
    *** Your goal (important):
    1. Only use the information above (recipe, kitchen items, etc.), Analyze the state of the kitchen and items, as well as the recipe.
    2. Decompose the recipe needed to finish cooking the soup into subtasks, with its subtask type, status, and all possible target locations
    3. Arrange these subtasks in chronological order.
    In this turn, please:
    - Generate a a Directed Acyclic Graph (DAG) of subtasks in the correct chronological order.
    - Output these subtasks in the given structure:    

\end{lstlisting}

\textbf{Graph Revision With Dialogue} 
After the node graph is generated by \textit{Coordinator}, a conversation will be initiated for revising the node graph. 

\begin{lstlisting}
System:|
    You are a robot assistant that breaks down tasks into steps and checks with the user when you are unsure about any part. At each step, if you are uncertain or the information is incomplete, ask the user for clarification or confirmation before proceeding.
    
    You are now collaborates with human, and you are responsible for make the subtask graph for achieving the recipe, which contains tasks for both human and you, the graph should follow this form {subtasks_example} 
    human will query you with some lauguage instructions, for every human query, first return the coordination type as 
    1 if human wants to change the coordination graph that will hold for continous collaborating
    2 if human want to indicate their preference
    3 if you human want to assign temporary tasks
    you should return 0 if you are even a little bit uncertain, you should send a short message to explicitly ask which type.
User: | 
    {Human_message}
Assistant: | 
    Query type: {query_type}
    
Examples: |     
    #Query Type 1: 
    Instructions: 
        human want to change the graph to create a long coordinating strategy, you should update the graph and adding or inserting nodes based on needs
    Assistant: | 
        Message to Human: {Human_message}
        updated Graph: {Node_graph}
        
    #Query Type 2: 
    Instructions: 
        human report preferences for different tasks, you should add to notes for later task assignment,  keep in mind that you are the robot.
    Assistant: 
        Message to Human: {Human_message}
        updated Graph: {Node_graph}

    #Query Type 3: 
    Instructions: 
        human querys a temporary subtask, you should add a node with status emergency on the basis of original graph, dont change original graph add in notes who should execute, keep in mind that you are the robot
    Assistant: | 
        Message to Human: {Human_message}
        updated Graph: {Node_graph}

\end{lstlisting}

\textbf{Active Suggestion} 
During the gameplay, \textit{Coordinator} can actively suggest coordination strategies to the human player based on current states and the node graph. 
\begin{lstlisting}
System: |
    You are coordinator, based on the current subtasks node graph, you want to 
    1. find which edge have a huge cost
    2. find potential collaboration intermediate point, where by collaborating on this location, the cost would be reduced.
    3. Are there any tricks when assigning different subtasks to different agent that can avoid collision. 
Instructions:|
    The normal procedure to finish one soup is:
    1. Pick up the required ingredients 
    2. Put ingredients into pots.
    3. start cook the soup.
    4. Pick up a dish.
    5. pick up the soup after it is ready.
    6. Put the soup to the serve location.
    
    Remember, you must put the all ingredients and correct ingredients exactly as specified in the recipe book, a important thing that you have to 
    put the soup to the serve location
    Recipe book:
    {recipe_book}
    
    Kitchen state:
    {kitchen_items}
    
    Current graph: 
    {current_graph}
    
    Analyze the environment layout and the current generated graph, analyze if there existing path cost between each subtask node is too high and thus the human and robot can collaborate to finish together, 
    that will reduce the cost, e.g. the robot could put ingredient or dishes into a certain empty couter, please indicate locations and human can pick up it from here
    Or if there will be tricks/preferences when assigning subtasks, like always assign a certain group of subtasks into human or robot will reduce the overall cost, give specific plans. 
    Return the suggestio as a short sentence into two variables, coordinator_suggestion for the first type, and preference_suggestion for the second type.     
\end{lstlisting}
\subsection{Manager: Subtask Assignment and Status Update} \label{app:prompt-manager}
\textbf{Subtask Assignment} 
Based on the current state and node graph, \textit{Manager} will actively assign subtasks to players while communicating with the human teammate. 
\begin{lstlisting}
System:|
    You are a task assigner, and please assign subtasks using the following rules
    1. do not assigne same task to human and robot
    2. Assign emergency subtask first
    3. Prioritize assign subtask to robot, if the robot are free, do not wait human to finish
    4. always handle the high priority subtasks and then consider the cost. 
Instructions:|
    Robot current state
        {robot_state}
    
    Human current state:
        {human_state}
    
    To finish the recipe, two agent (robot and human) are following a subtask graph, below are subtasks that are ready to execute:
    {graph_state}
    
    Your goal: 
    one of human or robots are not executing a task, you now want to reassign the subtask id for both human and robots, note that human and robot can not do the same task together. 
    Remember, you should prioritize assigning task to robot even if human are executing a task. Also return a very short message to instruct human task. 
    
\end{lstlisting}

\textbf{Subtask Status Update} 
\textit{Manager} actively monitors the status of the subtasks that players are currently working on and determines if humans or robots have finished the current executing subtask.
\begin{lstlisting}
System: |
    You are a task status judger, and please judge if an unexecuted subtask has been finished.
    1. note that there will be multiple repeated subtasks; you must make judgment based on the status change before and after robot interaction
    2. if a subtask is finished, the object held in the robot and human should change,d or the pot state will changed
    3. only check the current executing tasks if they are assigned
    4. you should be smart, if other unexecuted subtaskbut not currently executed by robot or human has been finished, human might not following the assinged tasks. 
Instructions:|  
    Robot States: 
    {robot_prev_state}
    {robot_state}

    Human States:
    {human_prev_state}
    {human_state}
    
    Robot assigned task
    {robot_task}
    
    Human assigned task
    {human_task}
    
    all other unexecuted tasks
    {all_possible_tasks}
    
    Have robot or human successfully finished the assigned task, based on there states before and after the interact, you have to judge if the interact does finished the subtask
    If the assigned task is start cooking or cook, as long as the pot has already start, even not finished, the assigned task should be treated as finished since it it automatically count
    Return a list of finished subtasks id to the variable finished_subtask_ids

\end{lstlisting}

\end{document}